\documentclass[11pt,twocolumn]{article}
\usepackage{hyperref}
\pdfoutput=1

\begin{document}

\title{\bf LES of Floating Wind Farms (Video Number V058)}

\author{Di Yang$^1$, Charles Meneveau$^{1}$, Richard Walters$^2$, Miguel Valenciano$^2$, \\
Michael Stephens$^2$, Randall Hand$^2$, and Lian Shen$^{1}$ \\[2pt]
\scriptsize $^1$Johns Hopkins University, Baltimore, MD, USA
\\[-2pt]
\scriptsize $^2$Unclassified Data Analysis \& Assessment Center,
U.S.\ Army Engineering Research \& Development Center, Vicksburg,
MS, USA}

%\vspace{-50pt}

\date{}

%\homepage[]{Your web page}
%\thanks{}
%\altaffiliation{}
%\affiliation{Department of Civil Engineering, Johns Hopkins
%University, Baltimore, MD 21218, USA}

\maketitle

%\begin{center} {\bf \Large LES of Floating Wind Farms} \end{center}

%{\hspace{-10pt}}{Di Yang$^1$, Charles Meneveau$^{1}$, Richard Walters$^2$, Miguel Valenciano$^2$,
%Mike Stephens$^2$, Randall Hand$^2$, and Lian Shen$^{1}$}

%{\it \scriptsize \hspace{-10pt}$^1$Johns Hopkins University, Baltimore, Maryland, USA

%\hspace{-10pt}$^2$Data Analysis and Assessment Center, U.S.~Army Engineering Research And
%Development Center, Vicksburg, Mississippi, USA

%}

%\vspace{-10pt}

%\begin{abstract}
%\section*{\normalsize Abstract}
%\vspace{-7pt}

\thispagestyle{empty}

\begin{abstract}

\vspace{-8pt}

The fluid dynamics video No.V058 is introduced, with brief
descriptions of the numerical method used to generate the animation
data, explanation of what is shown in the movies, and the main
scientific findings obtained from this study.
\end{abstract}

\vspace{-7pt}
\section*{\normalsize Simulation approach}
\vspace{-8pt}

Offshore wind energy has become an important frontier of sustainable
energy research.  In this study, large-eddy simulation (LES) of wind
turbulence coupled with potential-flow simulation of ocean waves is
performed for floating wind farms for the first time.  The LES of
marine atmospheric boundary layer is performed on boundary-fitted
grid that follows the wave motion.  The nonlinear evolution of
wavefield is simulated using a high-order spectral method.  The wind
and wave motions are coupled in the simulation by matching the
kinematic and dynamic boundary conditions at the sea surface.  Large
wind farm is modeled as periodic wind turbine array, with the
six-degrees-of-freedom motion of floating turbines solved subject to
the wind and wave loads, and the effect of turbines on wind modeled
using an actuator disc method.  Details of the numerical methods and
their validations are provided in Refs.~[1-3].

\vspace{-7pt}
\section*{\normalsize Description of videos}
\vspace{-8pt}

The first and second animations show respectively the perspective and top views of the simulated
ocean waves. The wavefield satisfies a JONSWAP spectrum, with peak wavelength of $52.3 \,\mbox{m}$
and significant wave height of $1.5 \,\mbox{m}$.  In the videos, the peak wave propagates from left
to right with a phase speed of $9.0\,\mbox{m/s}$.

The third animation shows the coupled motions of wind turbulence and ocean waves.  Sea surface and
wind speed on two vertical planes are shown.  The wind speed is normalized by its mean value at
$1\,\mbox{km}$ above the sea surface, which equals to $15.0\,\mbox{m/s}$ in this case.

The last two animations show two cases of LES of floating wind
farms.  In addition to sea surface and wind speed, turbine wakes are
illustrated with vorticity.  The periodic wind turbine array
consists of $3\times3$ turbines to represent part of a large wind
farm. In the first case, the sea surface is covered by JONSWAP waves
with parameters given above.  In the second case, the JONSWAP waves
are mixed with a swell with wave amplitude of $3.7\,\mbox{m}$ and
wavelength of $233.3\,\mbox{m}$. In the first and second cases,
because the waves are respectively small and large, the turbine
motions are weak and strong, respectively.

\vspace{-7pt}
\section*{\normalsize Main scientific findings}
\vspace{-8pt}

Analysis of the wind and wave statistics and turbine performance
shows that: (i) For large floating wind farms, the energy extracted
by the wind turbines is provided mainly by turbulence-mediated
downward flux of kinetic energy from the atmosphere above,
consistent with previous studies on land-based wind farms.$^{[3]}$
(ii) The wind field is substantially influenced by the waves due to
the effect on sea surface roughness and wave-induced form drag.  As
a result, the energy extraction rate of the wind turbines is a
function of wave conditions.  (iii) The motion of floating turbines
also affects the turbine performance.  For future development of
offshore wind farms, it is important to study the interaction among
wind, waves, and turbine motions.

\vspace{8pt}

{\scriptsize

{\hspace{-10pt}}$^{1}${D.~Yang} \& {L.~Shen}, {\sl J.~Comput.\
Phys.} {\bf 230}, 5510 (2011).

\vspace{2pt}

{\hspace{-10pt}}$^{2}${Y.~Liu}, {D.~Yang}, {X.~Guo} \& {L.~Shen},
{\sl Phys.\ Fluids} {\bf 22}, 041704 (2010).

\vspace{2pt}

{\hspace{-10pt}}$^{3}${M.~Calaf}, {C.~Meneveau} \& {J.~Meyers}, {\sl
Phys.\ Fluids} {\bf 22}, 015110 (2010).

}

\end{document}